\documentclass[prl,twocolumn, amsmath,amssymb, showpacs]{revtex4}

\usepackage{graphicx}
\usepackage{dcolumn}
\usepackage{bm}

\begin{document}

\title{A Multi-Path Interferometer on an Atom Chip}
\author{J. Petrovic$^{1,2}$, I. Herrera$^1$, P. Lombardi$^{1,3,4}$ and F. S. Cataliotti$^{1,5}$}
\affiliation{1) European Laboratory for Nonlinear Spectroscopy (LENS),
Via Nello Carrara 1, 50019 Sesto F.no (FI), Italy\\
2) Vin\v{c}a Institute of Nuclear Sciences, PO Box 522, 11001 Belgrade, Serbia\\
3) Dipartimento di Fisica e Astronomia Universit\`{a} di
Firenze via Sansone 1, 50019 Sesto F.no (FI), Italy \\
4)Laboratoire Kastler Brossel, Universit\'{e} Pierre et Marie Curie, Ecole Normale Sup\'{e}rieure, CNRS, Case 74, 4 place Jussieu, 75252 Paris Cedex 05, France\\
5) Dipartimento di Energetica "Sergio Stecco" Universit\`{a} di
Firenze via S. Marta 3, 50139 Firenze, Italy}

\begin{abstract}
Cold-atom interferometry is a powerful tool for high-precision
measurements of the quantum properties of atoms, many-body interactions
and gravity. Further enhancement of sensitivity and reduction
of complexity of these devices are crucial conditions for success of
their applications. Here we introduce a multi-path interferometric
scheme that offers advances in both these aspects. It uses coherent coupling
between Bose-Einstein condensates in different Zeeman states
to generate an interferometric signal with sharp fringes. We realise
such an interferometer as a compact easy-to-use
atom-chip device and thus provide an alternative method for measurement of the
light-atom and surface-atom interactions.
\end{abstract}


\maketitle

The first demonstration of coherence of a Bose-Einstein condensate
(BEC) \cite{KetterleSci97} has led to dramatic advancements in atom
interferometry. Long coherence times and the localization in phase
space of cold-atom clouds and in particular of BECs enable high
precision interferometric measurements of the internal properties of
atoms, many-body effects and gravity \cite{CroninRMP09}. Some
notable examples are the determination of spin squeezing
\cite{OberthalerNat10,RiedelNat10}, fine structure constant
\cite{GuptaPRL02}, density correlations \cite{HarberPRA02}, local
gravitational acceleration \cite{PetersMetro01}, Newtonian
gravitational constant \cite{FixlerSci07} and rotation rate
\cite{GustavsonPRL97}.

In the classical limit the best achieved sensitivity of a
conventional 2-path interferometer is determined by the shot noise and hence scales as $1/\sqrt{N}$
with the number of atoms $N$. This is known as the standard quantum
limit (SQL). An improvement in sensitivity beyond the SQL can be achieved by entangling the input and performing
a collective non-local measurement at the output. The best possible
outcome in a lossless system is the sensitivity of $1/N$, known as
the Heisenberg limit \cite{GiovannettiSci04}. An advancement in this
direction has recently been demonstrated by employing nonlinear
atom-atom interactions to produce the entanglement and therefore
reduce the phase-measurement error of a Ramsey interferometer below
the SQL \cite{OberthalerNat10}. An alternative way to improve the
interferometer sensitivity is to increase the fringe slope by
increasing the number of paths M as in \cite{WeihsOL96, WeitzPRL96}.
It is paid by a decrease in the average number of atoms per path and
hence by a greater susceptibility to noise. If the scaling of slope
with M exceeds $\sqrt{M}$ scaling of the shot noise, the sensitivity
improves with the number of paths. Multi-path interferometry can
be also seen as a fringe narrowing mechanism that increases
measurement resolution \cite{WeitzPRL96, MitchellNat04}.

Several multi-path matter-wave interferometers have been proposed,
the first being an atom-beam interferometer based on the optical
pumping between Zeeman states \cite{WeitzPRL96}. Multi-path
interferometry with cold atoms has been based on the property of an
optical lattice to impair a controllable recoil momentum on atoms as in \cite{KasevichSci98, FattoriPRL08, ChuPRL08, NaegerlNJP10, GuptaPRL02}. While they offer numerous advantages such as a
large number of paths, easy control of the relative phase
accumulation rate and compatibility with techniques for control of
atom-atom interactions,
these interferometers crucially depend on alignment, high-resolution
imaging and require a sophisticated technology to make them compact
and eventually portable.

Here we present a multi-path cold-atom interferometer with distinct narrow fringes that is simple to use and fully merged with an atom chip. The sharpness of the interferometric fringes is due to the interference of atoms at different Zeeman levels populated by high-frequency Rabi pulses. A nearly perfect coherence of this transfer yields a pure output signal without background. The interferometer can be used to measure parameters of state-dependent interactions of atoms with external fields, whereby the simultaneous measurement of multiple fringes at the output allows for application of multi-parameter sensing schemes.

\begin{figure}[ht]
\includegraphics[width=5cm,angle=-90]{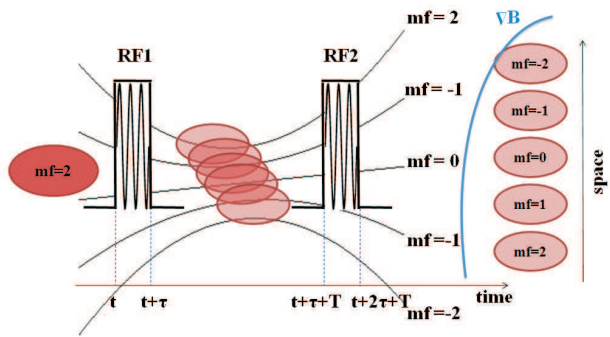}
\caption{Rabi pulses that comprise an integer number of RF oscillations act as controllable beam splitters on atoms in a BEC. Black curves show potentials of different Zeeman states in combined harmonic magnetic trap and gravity fields. The phase differences accumulated between the states are mapped into their populations at the output of the interferometer. A magnetic field gradient separates the states for imaging.}\label{Fig:Scheme}
\end{figure}

The essential components of the proposed scheme are shown in
Fig.~\ref{Fig:Scheme}. The initial state is prepared by condensing
atoms in a low field seeking ground state, here $|F=2,
m_F=2\rangle$. Coherent transfer of the atoms to other Zeeman states
of the same hyperfine state is realised by application of a resonant
RF pulse. The interferometer is closed by remixing these states by
the second RF pulse after a controllable time delay as in
\cite{MinardiPRL01}. The second pulse maps the relative phases
accumulated between different states during the delay into a
population distribution at the output of the interferometer. The
relative phases between the states are accumulated due to the
presence of the trapping magnetic field \textbf{B}. In this field
Zeeman states experience different potentials given by
$V=m_Fg_F\mu_0|B|$ where $m_F$ and $g_F$ are respectively the spin
and Land\`{e} numbers and $\mu_0$ is the Bohr magneton. Therefore,
their relative phases evolve with the frequencies equal to the
multiples of the energy separation between the adjacent levels
$\omega=g_F\mu_0 |\bf{B}|/\hbar$, yielding the interference signals rich in
harmonics. The harmonics cause the fringe width to decrease with the
number of states, which is the basic characteristic of a multi-path
interferometer. If an external signal is applied during the delay
between the pulses, it will contribute to the relative phases accumulated
between the states causing a shift in the fringe positions at the
output. Since the interferometric paths are not spatially separated,
the interferometer can be used to measure external fields whose
interactions with atoms are state-sensitive. Finally, in order to
determine the population of each output state, these states are
spatially separated by application of the Stern-Gerlach method
followed by the free-fall expansion and then imaged.

\begin{figure}[ht]
\includegraphics[width=6cm,angle=-90]{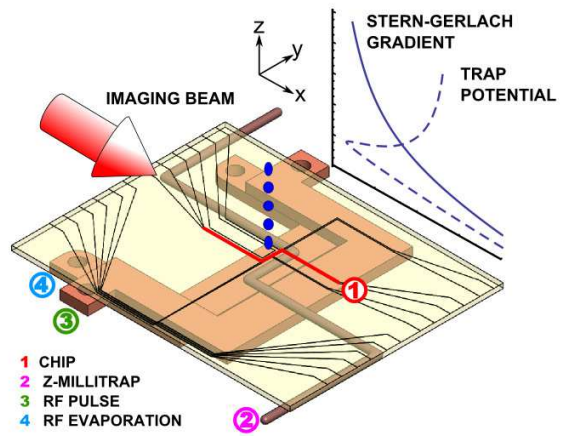}
\caption{The atom-chip interferometer. Magnetic trap on chip was formed by a micrometer z-wire (1)
and was loaded from ancillary magnetic trap formed by a millimeter Z-wire (2). Two U-wires served as RF antennas for Rabi pulses (3) and evaporative cooling (4). Plot shows z-wire magnetic fields with bias (z-trap) and without bias (Stern-Gerlach gradient). All Zeeman states were simultaneously detected by absorption imaging. The z axis points in the direction of gravity.}\label{Fig:Setup}
\end{figure}

Our interferometric setup was designed to be compact, fast and easy
to use. It comprises three essential parts: cooling apparatus,
RF-pulse source and detection system. A $^{87}$Rb BEC was realised on an atom
chip equipped with micrometer wires \cite{Vienna} and mounted on a
holder with additional millimeter U- and Z-wires as shown in
Fig.~\ref{Fig:Setup}. The condensate had 3x10$^4$ atoms and the critical
temperature of 0.5$\mu$K and was formed at 200$\mu$m from the chip
surface. It was kept in the cylindrically symmetric z-wire trap with
axial and transversal frequencies of 46Hz and 950Hz respectively
during the whole interferometric sequence. For each measurement a
new BEC was created.

The RF magnetic field was supplied by a waveform generator connected
to a U-wire in the chip holder. The pulse frequency matched the
$\sim1G$ separation of the Zeeman sublevels in the magnetic trap.
The pulse duration was short enough to ensure that the pulse
bandwidth covered the 10KHz energy spread of atoms in the BEC. We
note that the RF pulses were not locked in phase, but were phase
locked to their respective trigger signals as shown by the waveforms
in Fig.\ref{Fig:Scheme}.

The distribution of atoms across the Zeeman states was detected by a
7.5X-magnifying absorption imaging system with the imaging beam set to the $|F=2\rangle\rightarrow|F=3\rangle$ transition and a high-resolution camera. Prior to the imaging the Zeeman states were spatially
separated by the Stern-Gerlach method using a chip z-wire gradient.
The population of each state was normalized to the total atom number
making the interferometer independent of small fluctuations in the
condensate atom number.

The interferometer is well described by the transfer matrix $J$ that
acts on a vector of the complex amplitudes of states $\Psi(t)$ and
is given by the product $J=RPR$, where $R$ matrix describes the
M-level Rabi coupling and $P$ matrix the phase evolution of the
states. These two matrices are found as solutions of
time-dependent Schr\"{o}dinger equations
$i\hbar\frac{d\Psi(t)}{dt}=H\Psi(t)$ with the respective
Hamiltonians $H_{j,k}^R=\Omega/2(\delta_{j,j+1}+\delta_{j,j-1})$,
\cite{ShorePRA77}, and  $H_{j,k}^P=-i(j-1)\omega\delta_{j,k}$, where
$\Omega$ is the Rabi frequency and $\delta_{j,k}$ the Kronecker
symbol.
The R matrix elements are calculated by the technique proposed in
\cite{FujiiJMS08}.

The interferometer output has a form of a finite Fourier series
whose terms correspond to the multiples of the energy separation
$\omega$ between  adjacent Zeeman states in a magnetic field. An
increase in the number of paths M leads to an increase in the number
and order of the harmonics and hence to sharpening of the
interferometric fringes. We note that the transfer function is that
of a Fabry-Perot interferometer with a limited number of passes.
Moreover, the tridiagonal Rabi transfer matrix is analog to that of
an optical lattice \cite{TrombettoniPRL01} and an optical waveguide
array \cite{Yariv}. We note that for a number of states larger than
3 the eigenfrequencies of the Hamiltonian $H^R$ are incommensurate
to each other making it impossible to define a single pulse capable
of creating an arbitrary population distribution. One is therefore
limited to optimizing the pulse area for the best possible
performance.

\begin{figure}[ht]
\includegraphics [width=7cm,angle=-90] {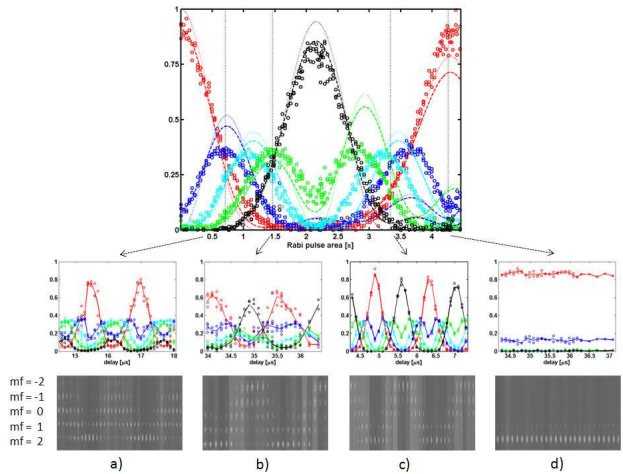}
\caption{Upper graph: Rabi pulse population transfer between $m_F=2$ (red), $m_F=1$
(dark blue), $m_F=0$ (light blue), $m_F=-1$ (green) and $m_F=-2$
(black) states as a function of the pulse area. Circles - experimental data, dotted line - theoretical result for the ideal initial conditions, dashed line -  theoretical result for experimental initial conditions.
Lower graphs: state populations at the output of the interferometer and the
corresponding absorption images for the RF-pulse areas of (a)
0.71$\pi$, (b) 1.46$\pi$, (c) 3.34$\pi$, and (d) 4.26$\pi$.}
\label{Fig:Rabi}
\end{figure}
An efficient use of the multiple Zeeman levels depends on the
availability of atoms in these levels between the two RF pulses and
therefore on the population transfer effected by the first Rabi
pulse. The role of the Rabi pulses is analogous to the role of beam
splitters in a Mach-Zehnder interferometer. However, here we can
control the splitting by varying the pulse parameters, the feature
not easily achievable in standard optical setups. The dependence of the
interferometer signal on the Rabi pulse area can be understood from
Fig.~\ref{Fig:Rabi}. The upper graph shows the measured and
calculated population transfers between all five Zeeman states by a
single Rabi pulse. Lower graphs show characteristic output signals
of the interferometer and the raw absorption images from which these
signals are derived. They correspond to the Rabi pulse areas marked
by the dashed lines. The Rabi pulse that populates only the highest
or the lowest Zeeman states (with $\Omega\tau$ around $0$, $2.1\pi$
or $4.2\pi$) yields interferometric fringes with low visibility
\cite{visibility} that approaches zero in the extreme case of the
complete population cycle, Fig.~\ref{Fig:Rabi}d). Well defined
narrow interferometric fringes with enhanced sensitivity can be
generated by the pulses that populate most Zeeman states as then
most harmonics allowed by the system contribute to the output
signal, see plots a)-c) in Fig.~\ref{Fig:Rabi}. The observed control
of the fringe sensitivity achieved by varying the Rabi pulse area is a time
analog of the control of interferometer output by a variable
beam splitter.

\begin{figure}[ht]
\includegraphics [width=6cm,angle=-90]{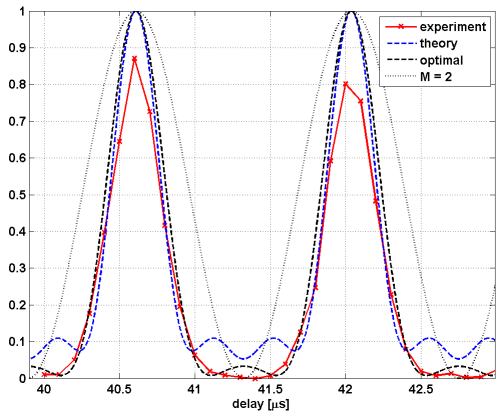}
\caption{Interferometric signals for the $m_F=2$ state: x - best experimental signal obtained for $\Omega\tau=3.34\pi$  (red line serves to guide the eye), blue dashed line - corresponding theoretical signal, black dashed line - optimal theoretical signal of a 5-path interferometer, black dotted line - optimal theoretical signal of a two-path interferometer. The realised interferometer has 1.75 times higher resolution than the best 2-path interferometer and approaches the resolution of the best 5-path interferometer.}
\label{Fig:Sens}
\end{figure}

We further used this feature to optimise the output of the
interferometer. The sharpest fringes with no side lobes are produced
by the Rabi pulses with the area of $3.34\pi$. Curiously, the
realised interferometer outperforms the theoretically predicted as
its signal is free from the background observed in the simulation
result in Fig.~\ref{Fig:Sens}. Instead, the measured approaches the
optimal theoretical signal of the 5-path interferometer also shown
in Fig.~\ref{Fig:Sens}. We attribute this discrepancy to the
simplifications of the model as well as to the experimental error. To evaluate the interferometer performance,
we compared the measured fringe slope with that of the ideal 2-path
interferometer. The optimal fringe slope of our interferometer is
$0.63\pm0.09$ rad$^{-1}$ and is larger than the 0.5 rad$^{-1}$ slope
of the 2-path interferometer. The improvement is further confirmed
by the 1.75 times fringe narrowing. Therefore, we have shown that
for an appropriate splitting of states, multiple paths interfere
constructively, which results in increase in the interferometer
resolution beyond that of an ideal 2-path interferometer.

However, we have not improved the sensitivity in the shot noise
limit. This is due to the slow increase in the fringe slope for
small M with respect to $\sqrt{M}$ scaling of noise. For M=5 the
ideal slope of $0.82$ rad$^{-1}$ allows for the realisation and
detection of an improvement only if the measurement error is below
3\%. Such precision was not reachable in our measurements. A more
favourable scaling could be achieved by allowing side lobes and aiming
for fringe densification, in analogy with the multi-photon states in
quantum optics \cite{MitchellNat04, WalmsleyNatPhot09}.

A remarkable feature of our interferometer is that the enhancement
of resolution is achieved without reduction in visibility. Indeed,
the complete transfer of atoms from $m_F=2$ to $m_F=-2$ and vice
versa confirms the coherence of the transfer and renders the fringe
visibility of 1 within the 5\% experimental error. By the
construction of the output signal, the high visibility is a
necessary but not sufficient condition for good performance of the
interferometer. For instance, the visibility of the $m_F=-2$ fringe
in Fig.~\ref{Fig:Rabi}a) is high but its sensitivity is low. The
maximum visibility is maintained for the delays between the pulses
of up to 50$\mu$s. For longer delays it decays exponentially with
the half-maximum time decay constant of 100$\mu$s.

Applications of the proposed interferometer are based on different
responses of the Zeeman states to an external field. Due to the
simultaneous measurement of multiple state populations, the
interferometer can be used in two basic measurement configurations:
absolute measurement in which the signal is defined as a shift of
fringes belonging to a chosen state, and differential measurement in
which the signal is defined as the difference in shifts of fringes
belonging to different states. An obvious application of the former
is in a magnetometer in which the magnetic field magnitude directly
maps into the periodicity of the fringes. Other applications are in
the measurement of parameters of light-atom interactions, for
example the relative atomic polarisability, and of light signals.

A potential disadvantage of the proposed sensor is its
cross-sensitivity to the magnetic field. This problem can be
resolved by using the differential measurement. In a magnetic field
with a changing magnitude all fringes experience the same shift that
cancels out in the differential signal. The remaining shift contains
only the desired information on the state-dependent interaction of
atoms with the external field. The elimination of the
cross-sensitivity is paid by a decrease in sensitivity and is thus
more efficient for a higher number of paths and well defined
fringes. We note that a strong interaction can distort the fringes
making the readout of the signals difficult. However, this does not
have severe consequences as it limits the application of the
interferometer exactly to the regime in which it performs best - the
small-signal regime in which the high sensitivity is required.

Finally, we observe that the simultaneous measurement of five output
signals enables the application of our interferometer as a
multi-parameter sensor. The multi-parameter sensing scheme is based
on the difference in sensitivity of Zeeman states to the same signal
and simultaneous sensitivity of each state to multiple signals. This
concept is well known in photonics where it is employed in
interrogation schemes with fibre gratings \cite{TatamMST03}, but it
has not been used in cold-atom based sensors before.

The integration of the interferometer with an atom chip offers several technical
advantages. The small wires allow for fast switching of the magnetic
fields thereby shortening the experimental cycle several times with
respect to the free-space setups. This is of particular importance
for the time-domain interferometers in which the signal is
constructed as a series of measurements at different time delays.
The proximity of the condensate to the chip wires enables large
field gradients that facilitate imaging. The chip interferometer is robust and
easy to use and as such is a good candidate for future portable cold-atom
sensors. Finally, the advantage that applies equally to the non-chip
setups is that the interferometer does not use light signals and
therefore does not suffer from the instabilities related to the
optical alignment.

To conclude, we have demonstrated a compact time-domain multi-path
interferometer on an atom chip whose sensitivity can be controlled
by an RF pulse acting as a variable beam splitter. In the optimal
configuration it has the resolution superior to that of an ideal
two-path interferometer. The simultaneous measurement of multiple
signals at the output enables a range of advanced sensing
applications in atomic physics and optics, while the integration of
interferometer with the chip puts it into consideration for future
portable cold-atom based measurement apparatuses. The presented
multi-path interferometric scheme can be applied to other atomic
species and any BEC setup, and is compatible with the quantum state
preparation. Moreover, the analogies of the couplers used here with
the optical lattices and coupled optical waveguide arrays indicate
the universality of the proposed scheme.

The authors wish to thank M. Inguscio for the continuous support and
many useful discussions. We wish to acknowledge the help of L.
Consolino and S. Bartalini in the construction of the atom chip
setup and useful discussions with C. Fort, M. Fattori and G. Roati.
We acknowledge the financial support of the Future and Emerging
Technologies (FET) programme within the Seventh Framework Programme
for Research of the European Commission, under FET-Open grants
MALICIA (265522) and CHIMONO (216774). J.P. acknowledges support of
the Ministry of Science of Serbia (Project III 45010). F.S.C.
acknowledges support of MIUR (Project HYTEQ).

\bibliographystyle{plain}

\begin{thebibliography}{99}

\bibitem{KetterleSci97} M. R. Andrews, C. G. Townsend, H.-J. Miesner, D. S. Durfee, D. M. Kurn, and W. Ketterle, Science \textbf{275}, 637 (1997).
\bibitem{CroninRMP09} A. D. Cronin, J. Schmiedmayer, and D. Pritchard, Rev. Mod Phys. \textbf{81}, 1051 (2009).
\bibitem{OberthalerNat10} C. Gross, T. Zibold, E. Nicklas, J. Est\`{e}ve, and M. Oberthaler, Nature \textbf{464}, 1165 (2010).
\bibitem{RiedelNat10}  M. F. Riedel, P. B\"{o}hi, Y. Li,  T. W. H\"{a}nsch, A. Sinatra, and P. Treutlein, Nature \textbf{464}, 1170 (2010).
\bibitem{GuptaPRL02} S. Gupta, K. Dieckmann, Z. Hadzibabic, and D. E. Pritchard, Phys. Rev. Lett. \textbf{89}, 140401 (2002).
\bibitem{HarberPRA02}  D. M. Harber, H. J. Lewandowski, J. M. McGuirk, and E. A. Cornell, Phys. Rev. A \textbf{66}, 053616 (2002).
\bibitem{PetersMetro01} A. Peters, K. Y. Chung, and S. Chu. Metrologia \textbf{38}, 25 (2001).
\bibitem{FixlerSci07}  J. B. Fixler, G. T. Foster, and J. M. McGuirk, and M. A. Kasevich, Science \textbf{315}, 74 (2007).
\bibitem{GustavsonPRL97} T. Gustavson, P. Bouyer, and M. Kasevich, Phys. Rev. Lett. 78, 2046 (1997).
\bibitem{GiovannettiSci04} V. Giovannetti, S. Lloyd, and L. Maccone. Science \textbf{306}, 1330 (2004).
\bibitem{WeihsOL96} G. Weihs, M. Reck, H. Weinfurter, and A. Zeilinger, Opt. Lett. \textbf{21}, 302 (1996).
\bibitem{WeitzPRL96} M. Weitz, T. Heupel, and T. W. H\"{a}nsch. Phys. Rev. Lett. \textbf{77}, 2356 (1996).
\bibitem{visibility} Visibility is defined as $V=(A_{max}-A_{min})/(A_{max}+A_{min})$ where $A_{min}$ and $A_{max}$ are the minimal and maximal output signal amplitudes, as in reference \cite{NaegerlNJP10}.
\bibitem{MitchellNat04} M. W. Mitchell, J. S. Lundeen, and A. M. Steinberg, Nature \textbf{429} 161 (2004).
\bibitem{KasevichSci98} B. P. Anderson and M. A. Kasevich, Science \textbf{282}, 1686 (1998).
\bibitem{FattoriPRL08} M. Fattori, C. D'Errico, G. Roati, M. Zaccanti, M. Jona-Lasinio, M. Modugno, M. Inguscio, and G. Modugno, Phys. Rev. Lett. \textbf{100}, 080405 (2008).
\bibitem{ChuPRL08} H. M\"{u}ller, S.-W. Chiow, Q. Long, S. Herrmann, and S. Chu, Phys. Rev. Lett. \textbf{100}, 180405 (2008).
\bibitem{NaegerlNJP10} M. Gustavsson, E. Haller, M. J. Mark, J. G. Danzl, R. Hart, A. J. Daley, \and
 H.-C. N\"{a}gerl, \emph{New J. Phys.} \textbf{12}, 065029 (2010).
\bibitem{MinardiPRL01} F. Minardi, C. Fort, P. Maddaloni, M Modugno, and M. Inguscio, Phys. Rev. Lett. \textbf{87}, 170401 (2001).
\bibitem{Vienna} The chip was supplied by the Atom Institute of the University of Vienna through CHIMONO collaboration.
\bibitem{ShorePRA77} B. W. Shore and J. Ackerhalt, Phys. Rev. A \textbf{15}, 1640 (1977).
\bibitem{FujiiJMS08} K. Fujii, J. Math. Sci. \textbf{153}, 57 (2008).
\bibitem{TrombettoniPRL01} A. Trombettoni and A. Smerzi, Phys. Rev. Lett. \textbf{86}, 2353 (2001).
\bibitem{Yariv} A. Yariv, \emph{Optical Electronics in Modern Communications}, 5th Ed. (Oxford Univeristy Press, New York, 1997).
\bibitem{WalmsleyNatPhot09} K. Banaszek, R. Demkowicz-Dobrza\'{n}ski, and I. A. Walmsley, Nature Photonics \textbf{3}, 673 (2009).
\bibitem{TatamMST03} S. W. James and R. P. Tatam, Meas. Sci. Techol. \textbf{14}, R49 (2003).

\end{thebibliography}

\end{document}